\documentstyle[11pt,psfig,aaspp4]{article}

\begin{document}

\title{Self-similar collapse and the structure of dark matter halos: 
A fluid approach }

\author{Kandaswamy Subramanian}

\affil{ National Centre for Radio Astrophysics,
Tata Institute of Fundamental Research, Poona University Campus,
Ganeshkhind, Pune 411 007, India.}

\begin{abstract}

We explore the dynamical restrictions on the structure of 
dark matter halos through a study of cosmological 
self-similar gravitational collapse solutions.
A fluid approach to the collisionless dynamics 
of dark matter is developed and the resulting
closed set of moment equations are solved numerically including
the effect of halo velocity dispersions
(both radial and tangential), for a range of spherically
averaged initial density profiles. 
Our results highlight the importance of tangential velocity dispersions 
to obtain density profiles shallower than $1/r^2$ in the core 
regions, and for retaining a memory of the initial
density profile, in self-similar collapse. 
For an isotropic core velocity dispersion 
only a partial memory of the initial density profile is
retained. If tangential velocity dispersions in the core 
are constrained to be less than the radial dispersion,
a cuspy core density profile shallower than $1/r$ cannot obtain, 
in self-similar collapse.

\end{abstract}

\section{Introduction}

In hierarchical clustering theories of structure formation, 
like the cold dark matter (CDM) models,
small mass clumps of dark matter form first and gather into larger 
and larger masses subsequently. The structure of these dark matter 
"halos", is likely to be related to how the halos formed, 
the initial spectrum of the density fluctuations and to the 
underlying cosmology. Several properties of galactic and cluster 
halos can be well constrained by observations.  So, if the 
matter distribution in dark halos
are fossils which do depend on some of the 
properties of structure formation models, like their initial
power spectrum, one would have a useful observational handle 
on these properties. It is therefore necessary to understand what determines 
the matter distribution (or density profiles)
of dark matter halos {\it ab initio}. 
This forms the motivation of the present paper and the 
companion paper by  Subramanian, Cen and Ostriker (SCO99, 1999).

Further, Navarro, Frenk and White (NFW) (1995, 96, 97)
have proposed from their N-body simulations, 
that dark matter halos in hierarchical 
clustering scenarios develop a universal density profile,
regardless of the scenario for structure formation or cosmology.
The NFW profile has an inner cuspy form with the density
$\rho \propto r^{-1}$ and an outer envelope of
the form $\rho \propto r^{-3}$. 
Several investigators have found that the NFW
profile provides a moderately good fit to numerical simulations
( Cole and Lacey 1996,
Tormen, Bouchet $\&$ White 1997, Huss, Jain $\&$ Steinmetz 1997, 1999,
Thomas {\it et al.}, 1998). Recently, though, high resolution 
simulations of cluster formation in a CDM model, 
by Moore {\it et al.} (1998), yielded a core 
density profile $\rho(r) \propto r^{-1.4}$, 
shallower than $r^{-2}$, but steeper than the $r^{-1}$ form 
preferred by NFW, consistent with the earlier high resolution
work of Xu (1995). (For smaller mass halos, Kravtsov {\it et al.}
(1998) find the core density profile to be shallower than the NFW form).
It is important to understand these results
as well on general theoretical grounds.

In a companion paper SCO99, we explore the possibility that 
a nested sequence of undigested cores in the center of a halo, 
which have survived the
inhomogeneous collapse to form larger and larger objects, determine 
halo structure in the inner regions.
For a flat universe with a power spectrum of density fluctuations
$P(k) \propto k^n$, scaling arguments then suggest that
the core density profile scales as,  
$\rho \propto r^{-\alpha}$ with $\alpha = \alpha_n = (9+3n)/(5+n)$.
But whether such a scaling law 
indeed obtains depends on the detailed dynamics.

Similarity solutions often provide a tractable, semi-analytic 
route to study time dependent dynamics in complicated physical systems.
Fillmore and Goldreich (FG, 1984) and Bertschinger (B85, 1985)
derived such solutions for describing the purely radial collapse of 
cold, collisionless matter in a perturbed Einstein-de Sitter universe.
These solutions need to be generalised to incorporate tangential
velocity dispersions, which as we  see below, turn out to be
crucial to understand density profiles shallower than $1/r^2$. 
Some general, analytical aspects of the 
similarity solutions incorporating tangential velocity dispersions
are outlined in the companion paper SCO99. 
In the present paper, we consider these self-similar 
collapse solutions in greater detail, by deriving and solving numerically 
the scaled moment equations for such a collapse, including the 
effect of velocity dispersions.

In the next section we formulate the self-similar
collapse problem and introduce a fluid approach for its solution,
recapitulating the corresponding discussion in SCO99.
In Section 3 we derive scaled moment equations describing the
collapse and discuss their numerical solution. 
Specific numerical examples of self-similar collapse 
are given in Section 4. These solutions include the
effect of halo velocity dispersions (both radial and tangential),
and consider a range of spherically averaged initial density profiles.
The final section discusses the results and presents our conclusions.

\section{ The self-similar solution }

We summarize in this section, some of the properties 
of the similarity solution, that can be derived by
analytic arguments. Although much of this
section is mostly a recapitulation of Section 3 of SCO99, we 
include it here to make the present paper as self-contained as 
possible, and also to set the framework for the 
detailed numerical work which follows.

Consider the collapse of a single spherically
symmetric density perturbation, in a flat background universe. 
Assume the initial density to be a power law in radius.
We expect to describe the dynamics through a self-similar solution. 
FG and B85 looked at the self-similar evolution 
by following the self-similar particle
trajectory. We adopt a different approach, by examining directly
the evolution of the phase space density.
During the course of this work we have learned that several
authors (Padmanabhan 1994, unpublished notes; Padmanabhan 1996a, Chieze,
Teyssier and Alimi 1997; Henriksen and Widrow 1997)
have also adopted this approach to the purely radial self-similar
collapse of FG and B85. We will emphasise and incorporate here also an
additional aspect, the distinctive role
of non-radial motions (velocity dispersions) in self-similar collapse.

The evolution of dark matter phase space 
density $f({\bf r}, {\bf v}, t)$ is governed by the Vlasov Equation,
\begin{equation}
{\partial f \over \partial t} + {\bf v}. {\partial f \over \partial{\bf r}}
+ {\bf a}. {\partial f \over \partial{\bf v}} = 0
\label{Vlasov}
\end{equation}
where ${\bf r}$ and ${\bf v} = \dot {\bf r}$ are the proper co-ordinate 
and velocity of the particles respectively. Also the acceleration 
${\bf a} = \dot {\bf v} = - {\bf \nabla }\Phi$, with
\begin{equation}
{\bf \nabla}^2\Phi = 4 \pi G \rho = 4 \pi G \int f d^3 {\bf v}
\label{pois}
\end{equation}
By direct substitution, it is easy to verify that these equations admit 
self similar solutions of the form 
\begin{equation}
f({\bf r}, {\bf v}, t) = k_2 k_1^{-3} t^{-q -2p} F( {{\bf r}\over k_1 t^p}, 
{{\bf v}\over k_1 t^q}) ; \qquad  p = q + 1
\label{scalf}
\end{equation}
where $k_1,k_2$ are constants which we will fix to
convenient values below.
We have used proper co-ordinates here
since the final equilibrium halo is most simply described in these
co-ordinates. (The same solution in co-moving co-ordinates is
given in Padmanabhan (1996a)). Defining a new set of co-ordinates 
${\bf y} = {\bf r}/(k_1t^p)$, ${\bf w} = {\bf v}/(k_1t^q)$ and a scaled
potential $\chi =k_1^{-2} t^{-2q}\Phi$, 
the scaled phase space density $F$ satisfies 
\begin{equation}
-(q + 2p) F - p {\bf y}. {\partial F \over \partial{\bf y}}
-q {\bf w}. {\partial F \over \partial{\bf w}}
+ {\bf w}. {\partial F \over \partial{\bf y}}
-{\bf \nabla}_{\bf y}\chi . {\partial F \over \partial{\bf w}} = 0 ;
\label{valsc}
\end{equation}
\begin{equation}
{\bf \nabla}_{\bf y}^2\chi = 4 \pi G k_2  \int F d^3 {\bf w} .
\label{potsc}
\end{equation}

Consider the evolution of a spherically symmetric density perturbation,
in a flat universe whose scale factor $a(t) \propto t^{2/3}$. 
For self similar evolution, the density is given by
\begin{equation}
\rho(r,t) = \int f d^3{\bf v} =
= k_2 t^{-2}\int F(y, {\bf w}) d^3{\bf w} \equiv k_2 t^{-2} \psi(y)
\label{densc}
\end{equation}
where we have defined $r = \vert {\bf r} \vert$, 
$y = \vert {\bf y} \vert$ and used the relation $p = q+1$. 
For the flat universe, 
the background matter density evolves as
$\rho_b(t) = 1/(6 \pi G t^2)$. So the density contrast 
$\rho(r,t)/\rho_b(t) = \psi(y)$, where we take $k_2 = 1/(6\pi G)$.

\subsection{ Linear and non-linear limits}

Let the initial excess density contrast averaged over a 
sphere of co-moving radius $x= r/a(t) \propto rt^{-2/3}$ be a power law 
$\bar\delta(x,t_i) \propto x^{-3\epsilon}$. 
Since $\rho/\rho_b$ is a function of $y$ alone, the $\bar\delta(x,t)$
will also be a function only of $y$.
Note that, in the linear regime, the excess density contrast
averaged over a {\it co-moving} sphere,
grows as the scale factor $a(t)$. So one can write
for the linear evolution of the spherical perturbation
\begin{equation}
\bar\delta(r,t)= \bar\delta_0 x^{-3\epsilon}t^{2/3} \propto \bar\delta_0 r^{-3\epsilon}t^{2/3 + 2\epsilon} \propto
\bar\delta_0 y^{-3\epsilon}t^{- 3\epsilon p + 2/3 + 2\epsilon} ,
\label{lincon}
\end{equation}
where we have substituted  $r \propto y t^p$.
This can be a function of $y$ alone, 
for a range of $t$ in the linear regime iff
$- 3\epsilon p + 2/3 + 2\epsilon = 0$, which gives
\begin{equation}
p = {2 + 6\epsilon \over 9\epsilon}
\label{adet}
\end{equation}
We see that once the initial density profile is specified the 
exponents $p,q$ of the self similar solution are completely determined.

Consider now what happens in the non-linear limit.
The zeroth moment of the Vlasov equation gives 
\begin{equation}
{\partial \rho \over \partial t} + {\bf \nabla}_{\bf r}.(\rho \bar{\bf v}) = 0
\label{contm}
\end{equation}
Here $\bar{\bf v} = <{\bf v}>$ is the mean velocity.
(Henceforth both $<>$ or a bar over a variable
denotes a normalised moment over $f$).
In regions which have had a large amount of 
shell crossings, it seems plausible to demand that the halo particles
have settled to nearly zero average infall velocity, 
that is $ \bar{v_r} \equiv 0$.
From (\ref{contm}) , we then have $(\partial \rho /
 \partial t) = 0$, and therefore, in the non-linear regime,
\begin{equation}
\rho(r,t) = Q(r) = Q(yt^{p}) = {1 \over 6 \pi G t^{2}} \psi(y).
\label{nonc}
\end{equation}
This functional equation has only power law solution,
because of the power law dependences on $t$. 
Substituting $Q(r) = q_0 r^{-\alpha}$ into Eq. (\ref{nonc}), 
and using $r \propto yt^p$,
we obtain $y^{-\alpha} t^{-p \alpha} \propto t^{-2} D(y)$. This can
only be satisfied for range of $t$ in the non-linear regime
provided $p\alpha = 2$. So, for an initial density profile 
with a power law slope $3\epsilon$, the power law slope of the
density in the non-linear regime is given by,
\begin{equation}
\alpha = {2 \over p} = {9\epsilon \over 3\epsilon + 1} .
\label{nonpow}
\end{equation}
This result has been obtained by following the self similar
particle trajectory, by B85 (for $\epsilon =1$), and 
FG for $2/3 \leq \epsilon < 1$. We see that it can
be simply obtained by just combining the self-similar
solution $f$ and the static core condition.
(Obtaining the B85/FG result in this way has been 
independently noted by Padmanabhan (private communication,
unpublished notes 1994)).

What should we choose for the value of $\epsilon$? For a power 
law $P(k) \propto k^n$,
the fractional density contrast averaged over a co-moving sphere of
radius $x$, is distributed as a Gaussian, with a variance
$\propto x^{-(3+n)/2}$.
This suggests  a "typical" spherically averaged initial
density law for a halo collapsing around a randomly placed point
of the form $\bar\delta(x,t_i) \propto x^{-(3+n)/2}$, or
$3\epsilon = (3 + n)/2$. Suppose we use this value of $\epsilon$ for
the initial density profile of a halo. Then
the halo density in the staic core regions will be 
$\rho(r,t) \propto r^{-\alpha}$, where, 
substituting $ 3\epsilon = (3 + n)/2$ in Eq. ( \ref{nonpow} ) 
\begin{equation}
\alpha = \alpha_n = { 9 + 3n \over 5 + n}
\label{aln}
\end{equation}
Remarkably, this is the same form that scaling laws suggest, 
for the core of a collapsed halo, assuming that the cores of
sequence of sub-halos are left undigested, during the formation
of the bigger halo (see SCO99).
(In a paper which appeared during the course of this work,
Syer $\&$ White (1998) motivate the same form, in the case when bigger halos
form by purely merger of smaller halos).
Note that for $n < 1$ the density law given by (\ref{aln}) 
is shallower than $1/r^2$.

FG also showed that a power law slope shallower than $1/r^2$,
cannot obtain for purely radial collapse, 
And that while the above form for $\alpha$ should obtain
for $2/3 \leq \epsilon < 1$, for $\epsilon < 2/3$, one goes to
the limiting value $\alpha = 2$. However, this is only true for
purely radial trajectories (cf. White and Zaritsky 1992; 
Sikvie, Tkachev and Wang 1997). We see below, 
by considering the higher moments of the Vlasov equation, that
$\alpha < 2$ can only obtain if the system has 
non-radial velocity dispersions. 

\subsection {Jeans and Energy equations} 

Suppose we multiply the Vlasov equation by the components of 
${\bf v}$ and integrate over all ${\bf v}$. 
Assume there is no mean rotation to the halo, that is
$\bar v_{\theta} = 0$ and $\bar v_{\phi} = 0$. Then we get
\begin{equation}
{\partial(\rho \bar v_r) \over \partial t} 
+{\partial(\rho \bar{v_r^2}) \over \partial r} 
+{\rho \over r} (2\bar{v_r^2} - \bar{v_{\theta}^2} - \bar{v_{\phi}^2})
+ {GM(r)\rho \over r^2} = 0
\label{radm}
\end{equation}
\begin{equation}
\bar{v_{\theta}^2}  = \bar{v_{\phi}^2}
\label{thetm}
\end{equation}
Here $M(r)$ is the mass contained in a sphere of radius $r$.

Let us consider again a static core with
$\bar v_r \equiv 0$. 
The Jeans equation gives 
two equations for the three unknown velocity
dispersions, even for a static core. 
To see if one can close the system SCO99 considered
the second moments of the Vlasov equation (the energy equations).
However these will involve the third moments, or the
peculiar velocity skewness. Some form of closure hypothesis
is needed in a fluid treatment of the Vlasov equation. 
For this we proceed as follows:
One can firstly assume that 
initially the tangential velocities have zero skewness.
Then in purely spherically symmetric
evolution they would not develop any skewness, that is 
$\bar{v_{\theta}^3} = \bar{v_{\phi}^3} =
< v_{\theta}v_{\phi}^2 > = 0$ for all times.
Also if the initial velocity ellipsoid had one
of its principle axis pointing radially, we do not expect this axis
to become misaligned in purely spherical evolution.
This means we can assume $< v_r v_{\theta}^2 > =
\bar{v_r} \bar{v_{\theta}^2 } $.
Under these assumptions, and taking the static core 
condition $\bar v_r = 0$, we get,
$(\partial(\rho \bar{v_{\theta}^2})/\partial t) = 0$
or $\rho \bar{v_{\theta}^2} = K(r)$ independent of $t$.
For the self-similar solution we then have
\begin{equation}
\rho \bar{v_{\theta}^2} = K(r) = K(yt^p) = k_2k_1^2
t^{4q -2p}\int w_{\theta}^2 F(y,{\bf w})d^3{\bf w}
\label{tan}
\end{equation}
Once again substituting a power law solution $K(r) = K_0 r^s$,
to this functional equation, we get the constraint from matching
power of $t$ on both sides,
$ps =4q - 2p$. Using $p = q +1$,
we then get
$s = 2 - 4/p = 2 - 2\alpha$, and so
\begin{equation}
\rho \bar{v_{\theta}^2} = K_0 r^{2 - 2\alpha}
\label{tanvel}
\end{equation}

Integrating the radial momentum equation using
Eq. (\ref{radm}) , (\ref{thetm}), (\ref{tanvel}) and using 
$\rho = q_0 r^{-\alpha}$, we have 
\begin{eqnarray}
\bar{v_r^2} &=& r^{2 - \alpha} \left [ {K_0 \over (2 - \alpha) q_0} - 
{4\pi G q_0 \over 2(2-\alpha)(3-\alpha)} \right ] \nonumber\\
&\equiv& {1 \over (2 - \alpha)} \left [ \bar{v_{\theta}^2}(r) 
- {GM(r)\over 2r} \right ] .
\label{consist}
\end{eqnarray}
Several important points are to be noted from the
above equation\altaffilmark{1}. 
\altaffiltext{1}{A constant of integration could have arisen in 
integrating the Jeans equation over radius; however this 
is excluded as, for a static core, arguments similar to deriving 
(\ref{tan}) and (\ref{tanvel}),  give $\rho {\bar v}_r^2 \propto  
r^{2 - 2\alpha}$, for the self-similar solution.}
A crucial one is that, when $ \alpha < 2$, 
the RHS of  Eq. (\ref{consist}) can 
remain positive, provided one has a non zero tangential
velocity dispersions. 
If one has a purely
spherically symmetric collapse and zero tangential
velocities, then the density law cannot become shallower
than $\alpha=2$ and maintain a static core with 
$\bar{v_r}=0$.  This agrees with FG. 
Infact for any $\alpha < 2$, one
needs tangential velocity dispersions to be at least
as large as $GM/2r$, comparable to the gravitational potential
energy per unit mass.
Further, one can see that to obtain static cores with $\alpha < 1$, 
the required tangential dispersions have to be necessarily
larger than the radial velocity dispersions.
Also note that for $\alpha < 2$, all the components
of velocity dispersions decrease with decreasing radius,
as suggested by the simple scaling arguments of SCO99.

For a static core $\bar{v_r^2}$ should 
be independent of $t$. However suppose we look at the 
the energy equation for the radial velocity dispersion,
\begin{equation}
{\partial(\rho \bar{v_r^2}) \over \partial t} 
+{1 \over r^2}{\partial(\rho r^2\bar{v_r^3}) \over \partial r} -
{2\rho < v_r(v_{\theta}^2+v_{\phi}^2) >  \over r} 
+2 \bar{v_r} \rho {GM/r^2} = 0 .
\label{radsqm}
\end{equation}
This shows that, even when ${\bar v}_r = 0$, a time independent
radial velocity dispersion can only obtain 
if the radial velocity skewness $<(v_r -\bar{v_r})^3>$
is also zero. In the core regions where large amounts
of shell crossing has occurred, one can assume that 
a quasi "equilibrium" state obtains,
whereby all odd moments of the distribution function, over 
$({\bf v} - \bar{\bf v})$, may be neglected. 
Such a treatment will correspond to considering a fluid like 
limit to the Vlasov equation. 

However, the radial skewness
will become important near the radius, where infalling matter
meets the outermost re-expanding shell of matter. This region
will appear like a shock front in the fluid limit.
A possible treatment of the full problem in the fluid approach 
to the Vlasov equation then suggests itself. This is
to take the radial skewness to be zero both inside and outside a
"shock or caustic" radius, whose location is to be
determined as an eigenvalue, so as to match the inner core
solution that we determine in this section
with an outer spherical infall solution. 
One has to also match various quantities
across this "shock", using jump conditions,
derived from the equations themselves.
To do this requires numerical
solution of the self consistent set of moment
equations derived from the scaled Vlasov equation
(the main focus of this paper), to which we now turn.

\section{ Numerical solution of moment equations for self similar collapse}

\subsection{ Moment equations}

We write the scaled Vlasov equation (\ref{valsc}) in spherical co-ordinates
and take moments. Let us define $V=\bar{w_r}$,
$\Pi= < (w_r -\bar{w_r})^2>$ and $\Sigma=\bar{w_{\theta}^2} =
\bar{w_{\phi}^2}$. We also set the tangential velocity skewness
to zero. As explained above, we 
take the radial skewness to be zero both inside and outside a
"shock or caustic" radius. The shock location, say $y=y_s$ in scaled
co-ordinates, will be
determined as an eigenvalue, to the complete problem.
So we set $< (w_r - \bar{w_r})^3> = 0$ in the regions of
interest, $ y< y_s$ and $y > y_s$. 
The resulting moment equations can be further simplified
with a little algebra and then can be written in the following
more transparent form: 
\begin{equation}
{1 \over y^2}{d\over dy}\left[y^2 \psi V\right] -2\psi 
-py {d\psi\over dy} =0
\label{cont}
\end{equation}
\begin{equation}
(p-1)\psi V + \psi (V -py){dV\over dy} = -{1\over y^2}{d \over dy}
\left[y^2\psi \Pi\right] + {2\psi \Sigma \over y} - 
{\bar{M} \psi \over 6\pi y^2}
\label{euler}
\end{equation}
\begin{equation}
(V-py){d \over dy}\left[{\rm ln}\left({ \psi \Pi y^2 \over (\psi y^2)^3}
\right)\right] = 2p -2
\label{energy}
\end{equation}
\begin{equation}
(V-py){d \over dy}\left[\Sigma y^2\right] + (4p-2)\Sigma y^2 = 0
\label{tandisp}
\end{equation}
\begin{equation}
{d \bar{M} \over dy} = 4\pi y^2 \psi
\label{mass}
\end{equation}
These equations have obvious meaning: Eq. (\ref{cont}) is the 
continuity equation
for the scaled density, (\ref{euler}) the scaled Euler equation and 
(\ref{energy}) the scaled energy equation. Angular
momentum conservation is reflected in Eq. (\ref{tandisp}) for $\Sigma$.
It should be noted that the energy equation reflects the more
general conservation $P/\lambda^3$ along fluid trajectories,
where $\lambda = \rho r^2$ is an effective linear radial density
and $P = \lambda < (v_r -\bar{v_r})^2>$ is the effective radial
pressure. Infact our system behaves like a monodimensional
gas with an effective adiabatic index $\gamma=3$, provided 
one takes the density to be the linear radial density $\lambda$,
and defines the pressure $P$ as above.

Both radial and tangential velocity dispersions 
are likely to be generated during the inhomogeneous collapse to form the halo.
So it is natural take the initial, pre-collapse, velocity dispersions to be
small. In the problem we are treating, of purely spherical collapse,
the radial velocity dispersions will automatically be
generated when spherically collapsing shells start to cross,
that is where the radial skewness is
important. In our fluid approach we will be replacing this
region where radial skewness is important by a shock front.
On the other hand, note that there are no source terms in 
the moment equations for the tangential velocity dispersions. 
Indeed in a purely spherically symmetric problem
tangential velocities have to be necessarily introduced
in an ad-hoc fashion. They are only non zero if present in the
initial conditions. 

White and Zaritsky (1992) introduced tangential
velocities into their solutions 
by invoking a fictitious tangential force which
act on particles in each spherical shell, until the
shell turns around. Since in a general inhomogeneous 
collapse, one expects all the components of the velocity
dispersion to be generated together, the shock gives an alternate 
natural location to introduce a tangential velocity dispersion, as well.
We will do this here, for most of the numerical examples.
Further in a  non spherical,
asymmetric collapse, the random velocities induced at the
shock would be in general non-radial (cf. Ryden 1993), and the 
spherical models with both tangential and radial velocity
dispersions introduced at the shock, may represent this in a rough way.
(For comparison, we will also present a few examples in the next 
section, with the tangential
velocity dispersion introduced at the turn around radius.) 
Studies of halo formation using cosmological N-body
simulations, which are discussed in SCO99,
can redress this deficiency of the spherical treatment.

The evolution of the region before shell crossings is  
determined by the spherical infall solution. 
At some initial time $t_i$, let the excess density contrast
averaged over a sphere of proper initial radius $r_i$ be 
$\bar\delta_i(r_i)= \delta_0(r_i/r_0)^{-3\epsilon}$. Then the shell
initially at $r_i$ will turnaround and collapse when it has
expanded to a radius $r_i/\bar\delta_i(r_i)$ at a time $t=(3\pi/4)
t_i/\bar\delta_i^{3/2}(r_i)$. The radius of a shell turning around at
any time $t$ is given by $r_t(t) = r_{0t}(t/t_{0t})^p$ where
$p = (2 +6\epsilon)/9\epsilon$ 
is as in Eq. (\ref{adet}) . Also
$r_{0t} = (r_0/\delta_0)$ and $t_{0t}=(3\pi/4)t_i/\delta_0^{3/2}$,
are the turn-around radius and time of the shell initially at $r_0$.
Since $y=r/(k_1 t^p)$, a natural way of fixing the constant
$k_1$ is by taking $k_1 t^p = r_t(t)$. We will do this in what follows. 
Then turn around occurs at the scaled co-ordinate $y=1$.

A straightforward application of the spherical model 
(cf. Peebles 1980,, Padmanabhan and Subramanian 1992, 
Kumar, Padmanabhan, Subramanian 1995) then gives the solution
of the moment equations, when $\Pi = \Sigma=0$, 
in the region $y > y_s$. Expressed in  a parametric form we have 
for $y > y_s$,
\begin{eqnarray}
y &=& {r\over r_t(t)} = {(1-\cos\theta) \pi^p \over 2(\theta -\sin\theta)^p} ;
\qquad
V(y) = y {\sin\theta (\theta - \sin\theta) \over (1 - \cos\theta)^2} 
\nonumber\\
\psi(y) &=& {9(\theta - \sin\theta)^2 \over 2(1-\cos\theta)^3}
\left[ 1 + 3\epsilon -  {9\epsilon \sin\theta(\theta - \sin\theta)
\over 2 ( 1 - \cos\theta)^2} \right]^{-1} ; \qquad
\bar M(y) = {4\pi y^3 \over 3}
{9(\theta - \sin\theta)^2 \over 2(1-\cos\theta)^3}
\label{outsol}
\end{eqnarray}
This goes over to the standard growing mode solution in the
linear limit as $y \to \infty$.

\subsection{ Matching and boundary conditions}

The equations (\ref{outsol}) evaluated at $y=y_s$ gives the pre-shock
boundary conditions to the moment equations. 
To match the spherical infall solution to the core solution
for $y<y_s$ determined in the previous section, 
we have to specify the jump conditions
across the shock at $y=y_s$. These conditions can be derived
again in a straightforward manner from the moment equations.
Suppose we denote
the pre-shock values with subscript $1$ and post shock
values of all quantities with a subscript $2$. Also 
we wish to consider the case when the pre-shock $\Pi_1=0$. Then 
the scaled jump conditions are given by
\begin{equation}
\psi_2 = 2 \psi_1 , \qquad V_2 = py_s + {1\over 2} (V_1 - py_s), \qquad
\Pi_2 = {(V_1 - py_s)^2 \over 4}, \qquad \bar{M}_2 = \bar{M}_1
\label{jump}
\end{equation} 
Infact the jump conditions corresponds to taking $\gamma=3$ in the usual
fluid Rankine-Huginot jump relations.
These together with a non zero, {\it arbitrary} $\Sigma_2$, gives
the starting values for the numerical integration of 
the scaled moment equations (\ref{cont}) - (\ref{tandisp}) inward from
the shock location $y=y_s$. The eigenvalue $y_s$ is determined by requiring
the solutions to satisfy the inner boundary conditions
\begin{equation}
V = M = 0, \qquad  y=0
\label{bcs}
\end{equation}
To ensure the vanishing of the mass at $y=0$, we have in fact
integrated the scaled continuity equation and expressed
the scaled mass interms of the density and velocities. We have using 
(\ref{cont}) and (\ref{mass})
\begin{equation}
\bar{M}(y) ={ 4\pi y^2 \psi (V-py)\over 2 - 3p}
\label{melim}
\end{equation}
The scaled density for all $\alpha$ 
and the scaled dispersions for $\alpha > 2$,
are expected to be singular at the origin for the shocked 
infall solutions. So we scale out the
expected asymptotic behaviour, at $y\to 0$, before numerical integration.
If we are to obtain a  nearly static core, we expect
$V \to 0$ and $dV/dy \to 0$ as $y \to 0$.
In this case, an analysis of the moment
equation shows (see also section 2),
\begin{equation}
\psi(y) = y^{-\alpha} \tilde{\psi}(y), \qquad 
\Sigma(y)=y^{2-\alpha}\tilde{\Sigma}(y), \qquad
\Pi(y)=y^{2-\alpha} \tilde{\Pi}(y)
\label{redefn}
\end{equation}
where $\tilde{\psi}(y)$, $\tilde{\Sigma}(y)$ and $\tilde{\Pi}(y)$
are expected to tend towards a constant value as $y \to 0$.
The exact asymptotic dependence of $V(y)$ of course has
to be determined by the numerical solution.

The moment equations 
(\ref{cont}) - (\ref{tandisp}) are numerically integrated,
after eliminating the scaled mass using 
(\ref{melim}) and transforming to the dependent variables defined 
in Eq. (\ref{redefn}) . We adapted a NAG library routine which integrates
the differential equations  using 
a Runge-Kutta-Merson method, and solves the
boundary value problem with Newton iteration in a shooting and
matching technique. For a given $\epsilon$, and a sufficiently
large $\Sigma_2$ (when $\alpha < 2$), 
a unique value of $y_s$ is found to satisfy the inner boundary 
conditions of (\ref{bcs}). The moment equations lead to
two conservation laws which can be used to provide
a check on the numerical integration. These can be derived by using
Eq. (\ref{melim}) and the moment equations. We have
\begin{equation}
{ \tilde{\Sigma}(y) \over \tilde{M}^{\kappa}(y) } = {\rm const}, \qquad 
\kappa = { 4 -\alpha \over 3 - \alpha }
\label{angcon}
\end{equation}
representing angular momentum conservation, where 
$\bar{M}(y)= y^{3-\alpha} \tilde{M}(y)$.
And
\begin{equation}
{ \tilde{\Pi}(y) \tilde{M}^{\mu}(y) \over \tilde{\psi}^2(y) } = {\rm const},
\qquad  \mu= {2 - \alpha \over 3 - \alpha}
\label{encon}
\end{equation}
representing energy conservation.
At each point these two integrals of motion
were checked and had relative errors less than $ 10^{-8} - 10^{-9}$, 
as in B85. A possible additional constraint on a solution 
is the asymptotic condition 
given by Eq. (\ref{consist}), for an almost static core. 
In terms of the scaled variables,
static core solutions satisfy the constraint
\begin{equation}
\tilde{\Pi}(0) = {1 \over (2 - \alpha)} \left [ \tilde{\Sigma}(0) 
- {\tilde{\psi}(0) \over 3(3-\alpha)} \right ] .
\label{saconsist}
\end{equation}
The equations of course cannot be integrated in practice 
upto $y \equiv 0$, but only upto some small $y=y_m $ which
is generally $\sim 10^{-2.5} - 10^{-4}$. So this constraint can
be checked at this minimum $y$. Also, in practice,
$V(y_m)$ is not expected to be identically
zero (unlike $V(0)$), and so one has to set it to a very small but
non-zero value to obtain converged solutions. In general for the solutions
obtained here, $V(y_m)= V_m \sim -1.0 \times 10^{-6}$ to 
$-1.0 \times 10^{-3}$ and $V_m/y_m << 1$. 
And the constraint (\ref{saconsist}) is satisfied
at a few percent level. Let us now discuss particular
numerical examples.

\section{ Numerical examples}

\subsection{ Collapse onto a localised, overdense perturbation,
 $\epsilon = 1$ case}

First we look at self-similar spherical secondary infall onto an
initially localised, overdense perturbation,
by adopting $\epsilon=1$ and $\Sigma_2=0$. 
This problem was solved by B85 and FG by examining the
self similar particle trajectory.
The parameters for the numerical
solution obtained here with the fluid approach
are summarized in Table 1. We give there the assumed
parameters, the derived eigenvalue $y_s$ and the value of 
the scaled dependent variables at $y_s$ and at the minimum $y=y_m$. 
We find the eigenvalue $y_s = 0.4628$ for the above parameters.
B85 solving the problem by looking at particle trajectories
got the location of the outermost caustic as $y_s=0.364$.
The difference between the location of the shock as determined by
this work and B85 could be because we have replaced
a smooth transition region for the collisionless
fluid, where velocity skewness is important, by a discontinuous shock.
B85 found that the scaled density could be fitted
asymptotically by a form $\psi(y) \approx 2.79 y^{-9/4}$ when they
adopted a minimum $y = y_m \sim 0.02$ for the
particle trajectory. We can integrate our equations and
get converged solutions satisfying the boundary conditions
upto $y_m \sim 2 \times 10^{-4}$. 
We find that $\psi(y) =\tilde{\psi} y^{-9/4} \approx 3.1 y^{-9/4}$
at $y_m$, while at
$y \sim 2 \times 10^{-2}$ we find $\tilde{\psi} \approx 2.5$. These
numbers bracket the asymptotic value of $\tilde{\psi} \sim 2.79 $ 
obtained in B85. 
So there is reasonable agreement between our work and B85, given
the differences in the value of $y_m$ and
the very different approaches. 

In Figure 1 we plot $V(y)$, $log(\psi(y))$, $log(\Pi(y))$, against
$log(y)$ for this solution with $\epsilon=1$, $\Sigma_2=0$.
We can define the scaled
rotational velocity $U(y) = [ (GM(r)/r)/(r_t/t)^2 ]^{1/2}$.
In Figure 1 we also show a plot of $log(U^2)$ versus $log(y)$.
We see that the velocity, $V(y)$, smoothly tends to zero
as $y \to 0$. $V_m$ for this case was $-4.0 \times 10^{-5}$.
To compare the asymptotic dependence of the
scaled density, with that predicted above for a static core, we also show
in the $log(\psi)$ vs $log(y)$ plot, 
density laws $\psi(y) \propto y^{-\alpha}$,
with $\alpha=9\epsilon/(1+3\epsilon)$ predicted in section 2, 
(dashed line) and $\psi(y) \propto y^{-2}$ (dot-dashed-dot line). 
These are normalized to agree with $\psi(y)$ at the minimum $y$
shown in the figure. We see from figure
2 that as $y\to 0$, the density does go over to
$\psi(y) \propto y^{-\alpha} \propto y^{-9/4}$, as expected for
a static core, with $\epsilon =1$. 
Overall, we recover the results of B85 and FG
reasonably well with our fluid approach to the problem.

\subsection{ $\epsilon < 2/3$ cases and the importance of
tangential dispersions}

We then considered solutions for initial density profiles
shallower than $r^{-2}$, or $\epsilon < 2/3$. For
such shallow density profiles, if the collapse
were purely radial, FG showed that the final density profile
approaches a $1/r^2$ form. We find, as expected, 
that the nature of the solutions in this case, 
depends on the ratio of tangential to radial velocity dispersions.
We illustrate this by considering two values of this ratio,
which bracket the expected behaviour.

In Figure 2 we show the solution for the case $\epsilon=0.4$,
$\tilde{\Sigma}_2 = \Sigma_2 y_s^{2 - \alpha}
= 0.94$. The detailed solution for this case
is given in Table 2. For this solution, the value 
of $y_s=0.4955$. We show both $log(\Pi(y)$ (solid line) and
$log(\Sigma(y)$ (dashed line) in the same plot, so that they can
be easily compared. From the figure or the
table one sees that tangential velocity dispersions 
for this solutions are everywhere larger than the radial
dispersions, by a factor $\sim 1.3$. (Or $(\Sigma/\Pi)^{1/2} \sim 1.3$).
For $\epsilon =0.4$,
and a static core, we expect the scaled
variables, to have the asymptotic behaviour given in
Eq. (\ref{redefn}), with $\alpha = 18/11$. We see from
comparing the solid and dashed lined, in the $\psi(y)- y$ plot
of Figure 2, that the density rises as $
\psi \propto y^{-\alpha} \propto y^{-18/11}$
to a very good approximation, throughout the core.
Also the velocity dispersions and rotation velocities
decrease with decreasing radius, as the analytic
theory of Section 2 (or Eq.(\ref{redefn})) predicts. Indeed,
from Table 2, we see that all the variables
$\tilde{\psi}(y)$, $\tilde{\Sigma}(y)$ and $\tilde{\Pi}(y)$
tend to constant values as $y \to 0$ to an
excellent approximation. This case illustrates that
it is possible to obtain solutions for $\epsilon < 2/3$, which
have $\alpha = 9\epsilon/(1+3\epsilon) < 2$, provided the tangential velocity dispersions are large enough. 

To illustrate the effect of decreasing tangential
velocity dispersions, we show in Figure 3, the
properties of a solution with $\epsilon=0.4$, $\tilde{\Sigma}_2 =0.65$.
The parameters for this solution,
are given in Table 1. The location of the shock is at
$y_s=0.3797$. We could get converged solutions with
$y_m=4.6 \times 10^{-3}$, and with the velocity 
$V_m =-1.75 \times 10^{-3}$. The core regions
are nearly static but not completely so. But the 
constraint given by Eq. (\ref{saconsist}) is
satisfied at the $2\%$ level. For this
case the radial velocity dispersions are everywhere 
larger than the tangential dispersions, by a factor 
$\sim 1.15$ as $y \to 0$. One sees a large difference
between this solution (Figure 3)
and the one obtained for larger tangential
velocity dispersion (Figure 2). First we see that when
radial dispersion dominates, the density profile
is closer to the $\psi \propto y^{-2}$ form than the
$\psi \propto y^{-\alpha}$ form, although neither
provides a good fit. Second the velocity dispersions
are reasonably constant with radius 
as $y \to 0$ limit, instead of decreasing
with decreasing radius. The rotation velocity is
also flatter. 

We also considered smaller values of $\epsilon$. 
Figure 4 gives a solution with $\epsilon=1/6$,
(corresponding to $\alpha=1$ or $n=-2$ in $\alpha_n$), 
and $\tilde{\Sigma}_2 = 2.2$ and some parameters for this
solution are given in Table 1. The eigenvalue $y_s=0.2584$ and 
$(\Sigma/\Pi)^{1/2} \sim 1.33$ in the core. Note that for
$\alpha=1$, the constraint equation (\ref{saconsist}) 
for a static core implies that 
$\tilde{\Sigma}(0) = \tilde{\Pi}(0) + \tilde{\psi}(0)/6 $. 
So $\tilde{\Sigma}(0)/\tilde{\Pi}(0) > 1$ for
any solution with a static core. 
The density profile
shows a reasonable correspondence with the 
asymptotic behaviour expected taking $\alpha =1$;
$\psi(y) \propto y^{-1}$. The velocity
dispersions and the rotation velocity
also decrease with decreasing $y$, but do so a
little less rapidly compared to the predicted $\Sigma
\propto \Pi \propto U^2 \propto y$ form. 

In general, we find that, for smaller values of $\epsilon \le 1/6$ 
(or $\alpha \le 1$),
while it possible to find static core solutions for
a sufficiently large $\Sigma/\Pi$ ratio, it becomes increasingly difficult
to do so (obtaining a small enough $V_m/y_m$ ratio)
as one lowers the ratio of $\Sigma/\Pi$, to even slightly
smaller values. We considered for example a case with 
$\epsilon=1/6$, $\tilde{\Sigma}_2 = 2$.
This turns out to have $\Sigma/\Pi \sim 1$,
but we found that we could only decrease $V_m/y_m$ to
a value of order unity and get a solution. 
We get $V_m \sim 1.5 \times 10^{-2}$ and
$dV/dy \sim 0.06$ at $y_m$; so even though the core is not strictly
static, the LHS of the scaled Euler equation (\ref{euler}) is 
much smaller than each of the individual force terms. These  
nearly cancel each other making the 
the core quasi-static. We give a plot of
all the variables for this solution in Figure 5,
and a summary of some parameters in Table 1. We see that
the shape of the density profile
for this solution is mid-way between the $r^{-\alpha}\propto r^{-1}$
form and $r^{-2}$ form. The velocity dispersions are decreasing 
with radius but not as rapidly as predicted for a truly
static core.

At this stage it is worthwhile to note the following:
Recall that the static core condition used 
to derive the asymptotic scaling properties
of the density and velocity dispersions, involves assuming not 
only ${\bar v}_r(0)=0$ (the boundary condition adopted above), 
but also that the radial velocity vanishes
for a range of radii near the origin. This situation can strictly
obtain only if particles with a given turn around 
radius have a {\it minimum} radius of approach to the centre; 
so that the core at any radius $r$ is evacuated of particles 
having turn around radii larger than say, $R_t(r)$. 
Such an "evacuated" core will inturn obtain only 
if the distribution of angular momentum
has a "hole" near the origin of $(v_{\theta}, v_{\phi})$ plane.
Such an angular momentum distribution is indeed assumed 
(and relevant) in the work of White and Zaritsky (1992).
However, in the present work we are making the
statistical assumption that the
distribution of tangential velocities is well described by
its second moment (viz. the tangential dispersion), thereby
excluding distribution functions, which have a hole. This
assumption is quite reasonable for halo cores with are
forming by a general inhomogeneous collapse. However,
in this case, for any shell of particles which pass the caustic
at some epoch, there are always some particles with sufficiently small
angular momentum, that can approach close to the halo
core. So the halo core will not be strictly static,
a feature which will be more and more noticeable, as one
decreases tangential velocity dispersions relative to
the radial dispersions. This may account for our result that
(for $\epsilon < 2/3$), as one decreases $\Sigma/\Pi$, 
the density profile is steeper 
than the $\psi \propto y^{-\alpha}$ form expected for a 
strictly static core.

Finally, for the sake of comparison, we have also looked
at numerical examples where the tangential velocity dispersions
are introduced at the 'turn around' radius (taken to be
approximately at $y=1$) rather than at the
shock. In this case one has to solve the moment equations
numerically, both outside and inside the shock radius, match the
solutions across the shock, using the shock jump conditions 
(cf. Eq. (\ref{jump}) when $\Pi_1 = 0$), and find the shock 
location as an eigenvalue to satisfy the boundary conditions
in Eq. (\ref{bcs}). Figures 6 and 7 give two examples with 
$\epsilon=0.4$, adopting $\tilde{\Pi}(1) =0$ 
and $\tilde{\Sigma}(1) =0.25$ and 
$\tilde{\Sigma}(1) =0.30$, respectively. The parameters of
these solutions are given in Table 1. When 
$\tilde{\Sigma}(1) =0.25$, the force due to the tangential
velocity dispersion at turn around is $\sim 13.5 \%$ of the
radial gravitational force. These examples show very similar
behaviour to the $\epsilon = 0.4$ solutions discussed above 
(Figures 2 and 3), where the tangential velocities are 
introduced at the shock. For example, the solution shown
in Figure 6, has $\Sigma/\Pi \sim 1$ in the core. The
corresponding density profile is shallower than $y^{-2}$ 
but steeper than the $y^{-\alpha}$ form, reflecting the fact
that only a partial memory of the initial profile is
retained by self-similar evolution in this case.

The numerical results of this section shows the importance of
tangential velocity dispersions, in deciding whether the
self similar solution, with an initial density profile
shallower than $1/r^2$ ($\epsilon < 2/3$) retains a memory of this initial
profile or whether the density profile tends to a universal $1/r^2$ form.
The set of solutions we have given show that for
a large enough $\Sigma/\Pi > 1$, the
the core density profile is indeed close to the form 
$\rho \propto r^{-\alpha}$,
with $\alpha = 9\epsilon/(1+3\epsilon)$. For $\Sigma/\Pi \sim 1$, 
some memory of the
initial density profile is always retained; the density profile
has an asymptotic form $\rho \propto r^{-\bar\alpha}$, with
$ \alpha < \bar\alpha < 2$. 
When $\Sigma/\Pi << 1$, the density profile goes
over to the $1/r^2$ form derived by FG. Also for
shallow initial density profiles with $\alpha  \leq 1$,
one must necessarily have a tangential dispersion larger
than radial dispersion to get a static core region, 
retaining the memory of the initial density profile.

\section{Discussion and Conclusions}

We have explored here the dynamical restrictions on the structure of 
dark matter halos through a study of cosmological 
self-similar gravitational collapse solutions,
adopting a fluid approach to the collisionless dynamics 
of dark matter. In a companion paper (SCO99) we consider 
the possibility that a nested sequence of 
undigested cores in the center of a halo, which have survived the
inhomogeneous collapse to form larger and larger objects, determine 
halo structure in the inner regions.
For a flat universe with 
$P(k) \propto k^n$, scaling arguments then suggest that
the core density profile scales as,  
$\rho \propto r^{-\alpha}$ with $\alpha = \alpha_n = (9+3n)/(5+n)$.
However, such arguments do not tell us how and in fact whether this
form will be realized dynamically. The similarity solutions
worked out in some detail here, allows us to examine this dynamical
issue, in a simple tractable manner.

The problem of spherical self similar collapse,
has often been solved by following 
particle trajectories. We adopted here and in SCO99 another approach, 
examining directly the evolution of the moments of
the phase space density. For a purely radial
collapse, with the initial density profile $\propto r^{-3\epsilon}$, 
and steeper than $r^{-2}$, we recover, by demanding that the core be static,
the asymptotic form of the non-linear density profile: 
$\rho \propto r^{-\alpha} \propto r^{-9\epsilon/(1 + 3\epsilon)}$
(see also Padmanabhan 1996b). 
For initial density profiles shallower 
than $1/r^2$, with $\epsilon < 2/3$, we showed that, 
a static core with a non-linear density profile,
with $\alpha= 9\epsilon/(1 + 3\epsilon)$, is
possible, only if the core has sufficiently large tangential
velocity dispersions.  Infact, one
needs $\bar{v_{\theta}^2} > GM/2r$. 
Also if a static core has to have a cuspy density
profile shallower than $1/r$, (with $\alpha < 1$), 
one requires $\bar{v_{\theta}^2} > \bar{v_r^2}$. 
Note that when $3\epsilon = (3 +n)/2$ (as would be relevant 
for collapse around a typical point in the universe), 
$\alpha = \alpha_n = (9 + 3n)/(5+n)$.

The consequences of introducing non radial velocity dispersions,
in this approach, can only be examined in detail, by adopting
a closure approximation. 
In spherical collapse, the skewness of
the tangential velocities can be assumed to be zero,
in the core regions. In fact,
in regions where large amounts
of shell crossing has occurred, one can assume that 
a quasi "equilibrium" state obtains,
whereby all odd moments of the distribution function, over 
$({\bf v} - \bar{\bf v})$, may be neglected. 
The radial peculiar velocity is then
also expected to have negligible skewness, in the core regions.  
However, the
radial peculiar velocity will necessarily have a
non-zero skewness (non zero third moment) near a caustic radius,
where collapsing dark matter particles meet the
outermost shell of re-expanding matter. 
To take this into account we introduce a 
fluid approach. In this approach, the effect of peculiar 
velocity skewness is neglected in all regions except 
at location of the caustic,
which we call the shock. In the particle picture the shock
is where a single stream flow becomes a muti stream flow.
In the fluid picture it is a where some of the
average infall velocity, is converted to velocity dispersion. 
The location of the caustic, $y_s$, in
scaled co ordinates, is found 
as an eigenvalue, to the boundary value problem of matching 
the single stream collapse solution with 
a core solution, adopting $V=M=0$ as the 
boundary condition at $y=0$.

In spherical collapse tangential velocities are only non
zero if they are present in the initial condition. 
The shock or the turn around radius, provide 
a natural location for introducing tangential
dispersions, into the initial conditions. Our treatment here
assumes that the distribution of tangential velocities is
well described by just its second moment, consistent 
with the statistical assumptions of a quasi-relaxed core. 
The results of the numerical integration of the moment
equations, are summarized 
in Table 1 and are graphically displayed in Figures 1-7.
The details of one particular solution is also given in Table 2. 

These examples largely bear out the expectations of section 2.
First we recover quite well, using the fluid approach, the
the asymptotic form of the non-linear density profile, 
for the $\epsilon = 1$ case, which B85/FG got by solving for 
the self-similar particle trajectory. Second our solutions
show the importance of tangential velocity dispersions, 
in deciding the nature of the core density profile, 
when $\epsilon < 2/3$. In the spherical self
similar collapse solutions with $\epsilon < 2/3$, 
for a large enough $\Sigma/\Pi > 1$, 
one gets $\rho \propto r^{-\alpha}$,
with $\alpha = 9\epsilon/(1+3\epsilon)$. For $\Sigma/\Pi \sim 1$, 
some memory of the initial density profile is always retained; one gets
$\rho \propto r^{-\bar\alpha}$, with $ \alpha < \bar\alpha < 2$.
When $\Sigma/\Pi << 1$, the density profile goes
over to the $1/r^2$ form derived by FG for radial collapse. Also 
$\alpha  < 1$, requires $\Sigma/\Pi >> 1$,
to get a static core region. 
So if in halo cores tangential velocities are constrained
to be smaller than radial velocity dispersions, then a cuspy
core density profile shallower than $1/r$ cannot obtain,
purely by self-similar evolution.

The results of this work and SCO99,  illustrate
the importance of dynamical considerations and hint at
features which are likely to obtain in more realistic collapse.
If newly collapsing material is constrained to mostly contribute to
the density at larger and larger radii, then memory
of initial conditions can be retained.
The solutions, with $\alpha > 2$ (Figure 1), or 
with $\alpha < 2$ but a large enough tangential
dispersion (Figures 2 and 4), illustrate this possibility.
However  when newly collapsing
material is able to occupy similar regions as the matter
which collapsed earlier, the core density profile 
will only partially reflect a memory of the initial conditions.
The solutions in Section 4 with $\alpha < 2$ and $\Sigma/\Pi \sim 1$
(Figures 3, 5 and 6) illustrates this feature.

In SCO99 we have also adopted a complimentary approach, of
looking at halo properties in numerical simulations of
structure formation models having $n=-2,-1 $ and $0$.
We find that the core density profiles of dark matter halos
show a large scatter in their properties, but do nevertheless appear to
reflect a memory of the initial power spectrum (please see SCO99 for
details). The fluid approach adopted here and in SCO99 suggests 
new ways of exploring non linear dynamics. Perhaps one can 
extend analytic approximations like the Zeldovich
approximation, valid in a single stream flow, to
the multi streaming regime, by replacing 
multistreaming regions by regions with velocity dispersions, 
generated by the Zeldovich type caustics.
The fluid approach could also be useful 
to study possible closures of the BBJKY hierarchy. Further
one needs to extend the self-similar
solutions to incorporate a  baryonic component;
the gas necessarily has an isotropic
velocity dispersion, and so will have a different dynamical
evolution compared to the dark matter. We hope to study
some of these issues in the future.

\acknowledgments

KS thanks Jerry Ostriker and Renyue Cen for an enjoyable
collaboration which led to this paper.
This work was begun when KS visited the Princeton University
Observatory, during Sept-Nov 1996. 
Partial travel support to Princeton came from IAU Commission 38.
Some of the work was done at the University of Sussex where 
KS was supported by a PPARC Visiting Fellowship. 
He thanks John Barrow, Jerry Ostriker, Ed Turner, 
the other Princeton and Sussex astronomers for warm hospitality.
T. Padmanabhan is thanked for critical comments on
an earlier version of this work. KS also thanks
Ben Moore, Bepi Tormen, Ravi Sheth, Dave Syer
and Simon White for several helpful discussions.

\clearpage
 
%
%

\begin{figure*}
\centering
\begin{picture}(400,250)
\psfig{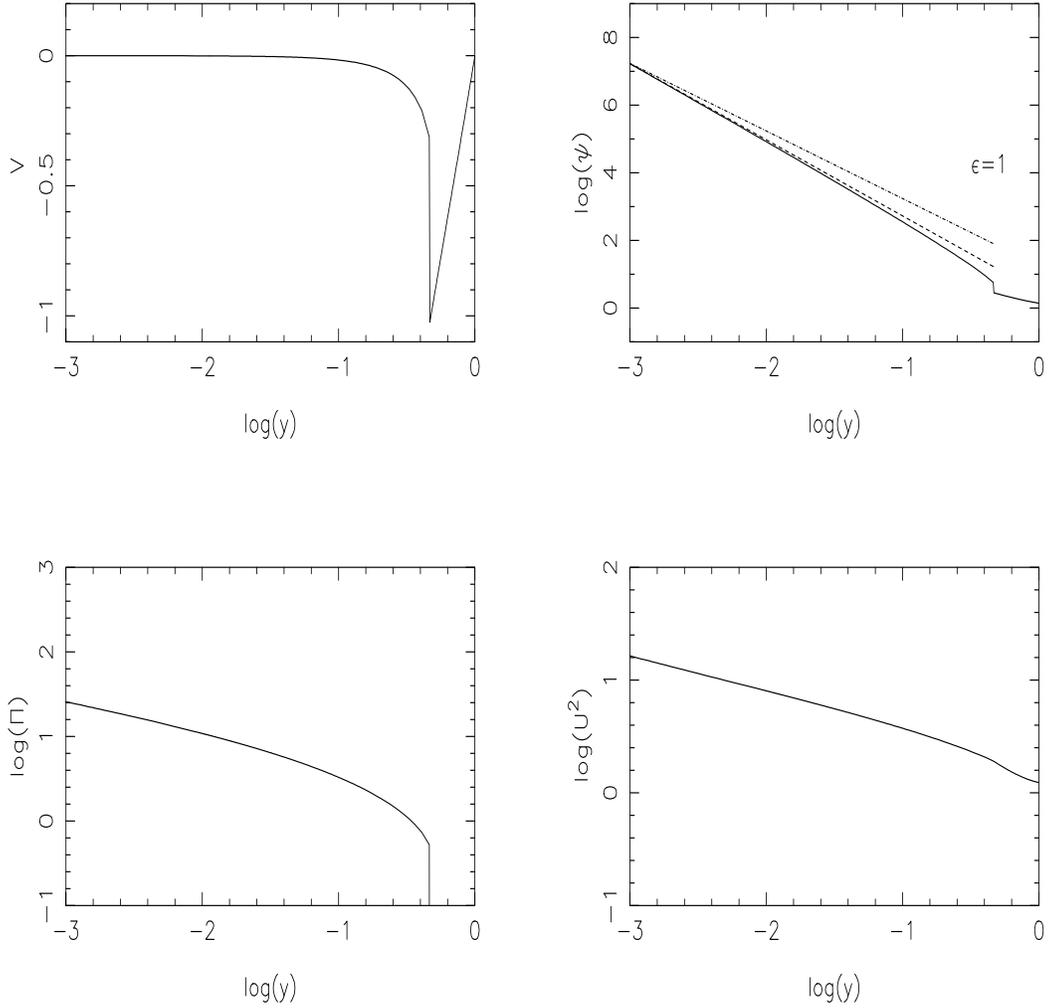}
\end{picture}
\caption{ Self similar collapse solution for $\epsilon=1$ case of B85/FG. 
 The velocity $V$, scaled density $\psi$ (solid line in the upper right plot), 
radial velocity dispersion squared $\Pi$, and circular velocity squared
$U^2$ are plotted against the scaled radius $y$. The pre shock 
spherical infall solution is also shown. In the $\psi$-$y$ plot
we also show for comparison the density laws $\psi \propto r^{-\alpha}$
(dashed line) with 
$\alpha=9\epsilon/(1+ 3\epsilon)$ 
and $\psi \propto r^{-2}$ (dashed -dotted line).}
\end{figure*}

\clearpage

\begin{figure*}
\centering
\begin{picture}(400,250)
\psfig{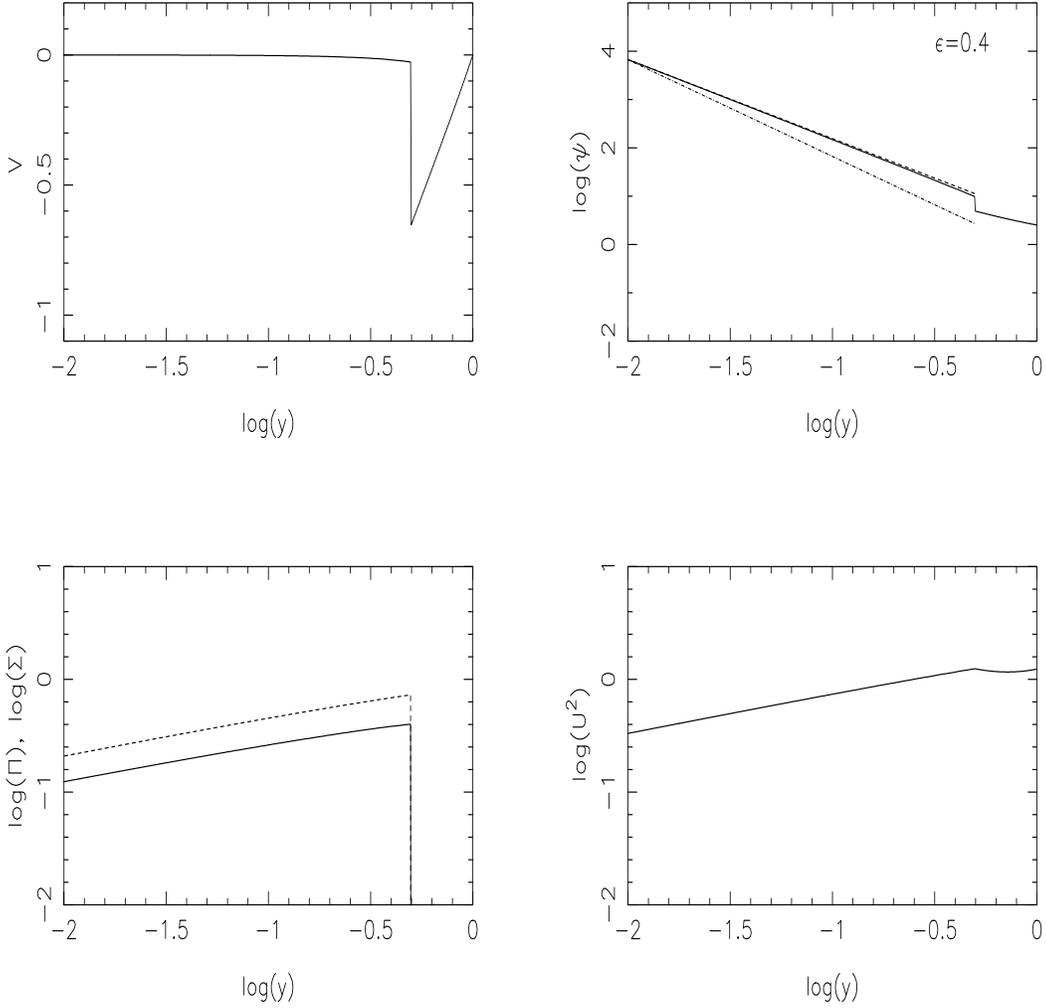}
\end{picture}
\caption{Self similar collapse solution for $\epsilon=0.4$, 
$\tilde{\Sigma}_2 = 0.94$.  
 The velocity $V$, scaled density $\psi$ (solid line in the upper right plot), 
radial velocity dispersion squared $\Pi$ (solid line in the lower left plot), 
tangential velocity dispersion squared 
$\Sigma$ (dashed line in the lower left plot) 
and circular velocity squared
$U^2$ are plotted against the scaled radius $y$. The pre shock 
spherical infall solution is also shown. In the $\psi$-$y$ plot
we also show for comparison the density laws $\psi \propto r^{-\alpha}$
(dashed line)  with 
$\alpha=9\epsilon/(1+ 3\epsilon)$
and $\psi \propto r^{-2}$ (dashed -dotted line).
}
\end{figure*}

\clearpage

\begin{figure*}
\centering
\begin{picture}(400,250)
\psfig{figure=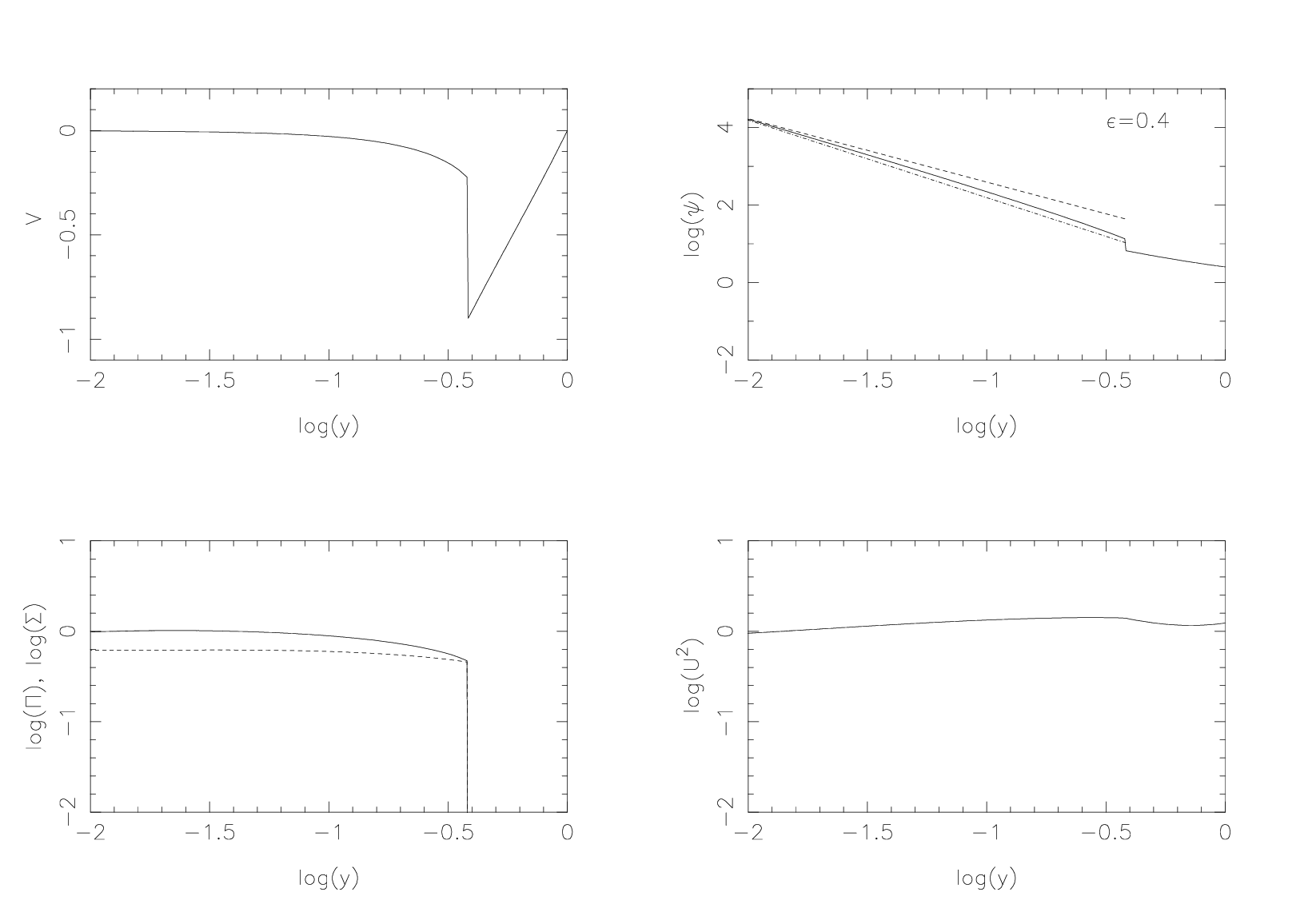,height=15.0cm,width=15.0cm,angle=-90.0}
\end{picture}
\caption{ Self similar collapse solution for $\epsilon=0.4$, 
$\tilde{\Sigma}_2 = 0.65$. The various quantities shown are same as
in Fig. 2.  
}
\end{figure*}

\clearpage

\begin{figure*}
\centering
\begin{picture}(400,250)
\psfig{figure=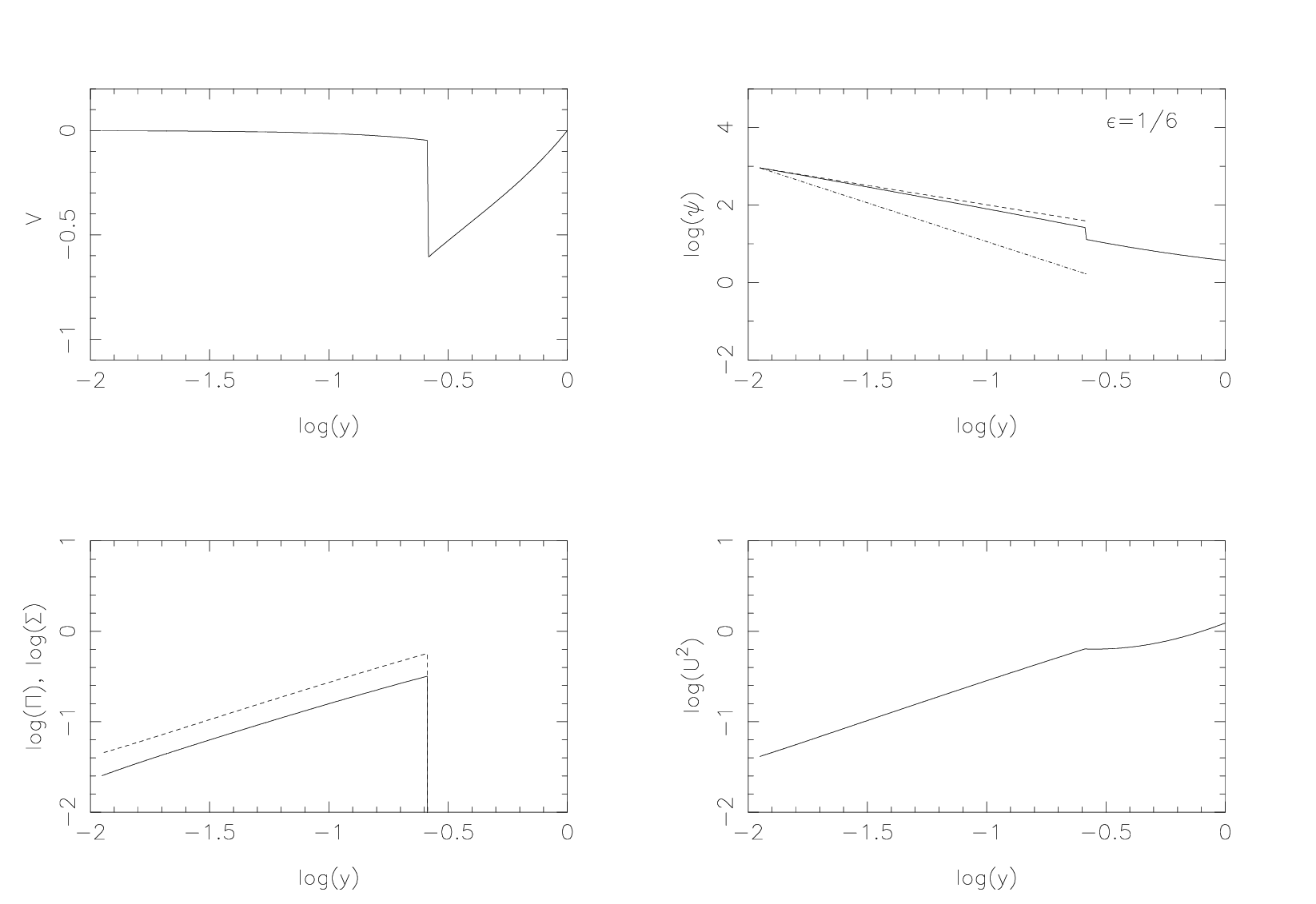,height=15.0cm,width=15.0cm,angle=-90.0}
\end{picture}
\caption{ Self similar collapse solution for $\epsilon=1/6$, 
$\tilde{\Sigma}_2 = 2.2$. The various quantities shown are same as
in Fig. 2. }
\end{figure*}

\clearpage

\begin{figure*}
\centering
\begin{picture}(400,250)
\psfig{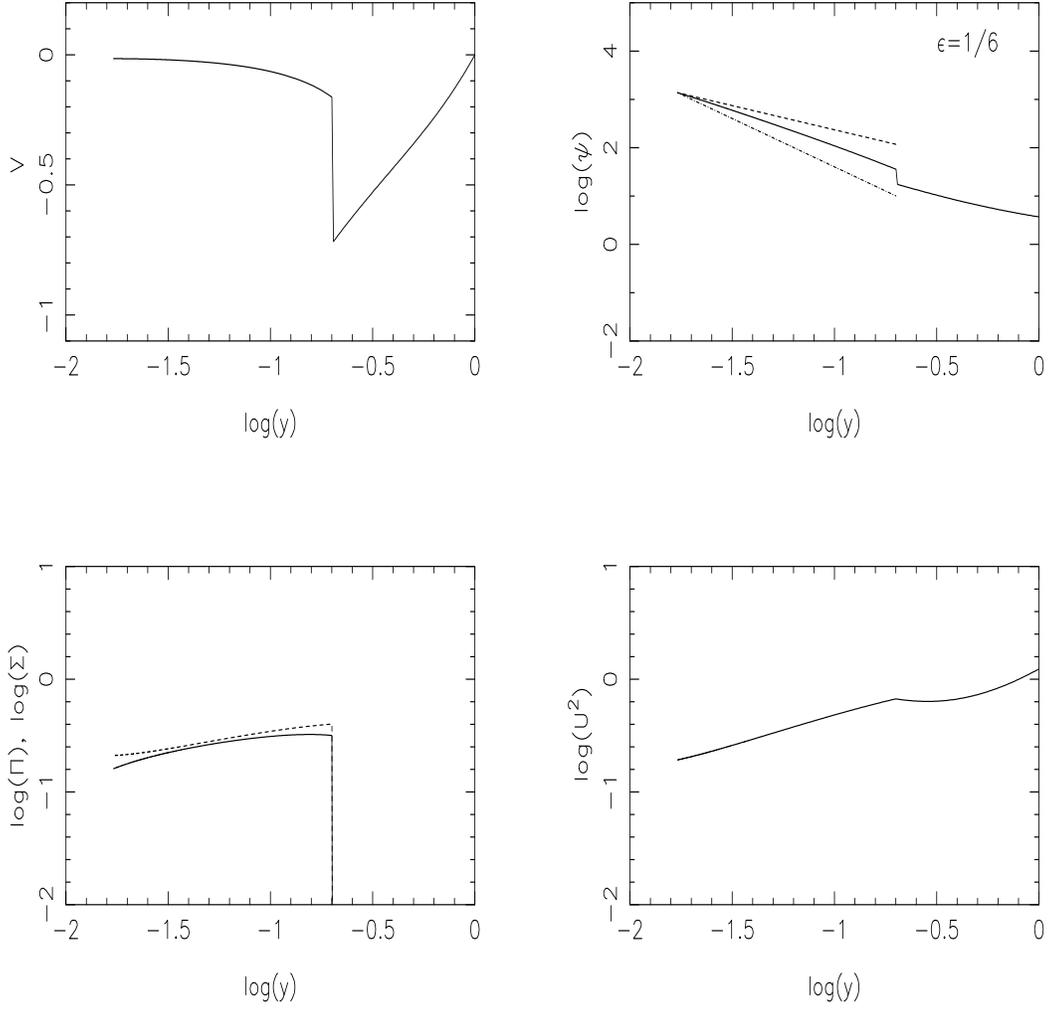}
\end{picture}
\caption{ Self similar collapse solution for $\epsilon=1/6$, 
$\tilde{\Sigma}_2 = 2.0$. The various quantities shown are same as
in Fig. 2.  }
\end{figure*}

\clearpage

\begin{figure*}
\centering
\begin{picture}(400,250)
\psfig{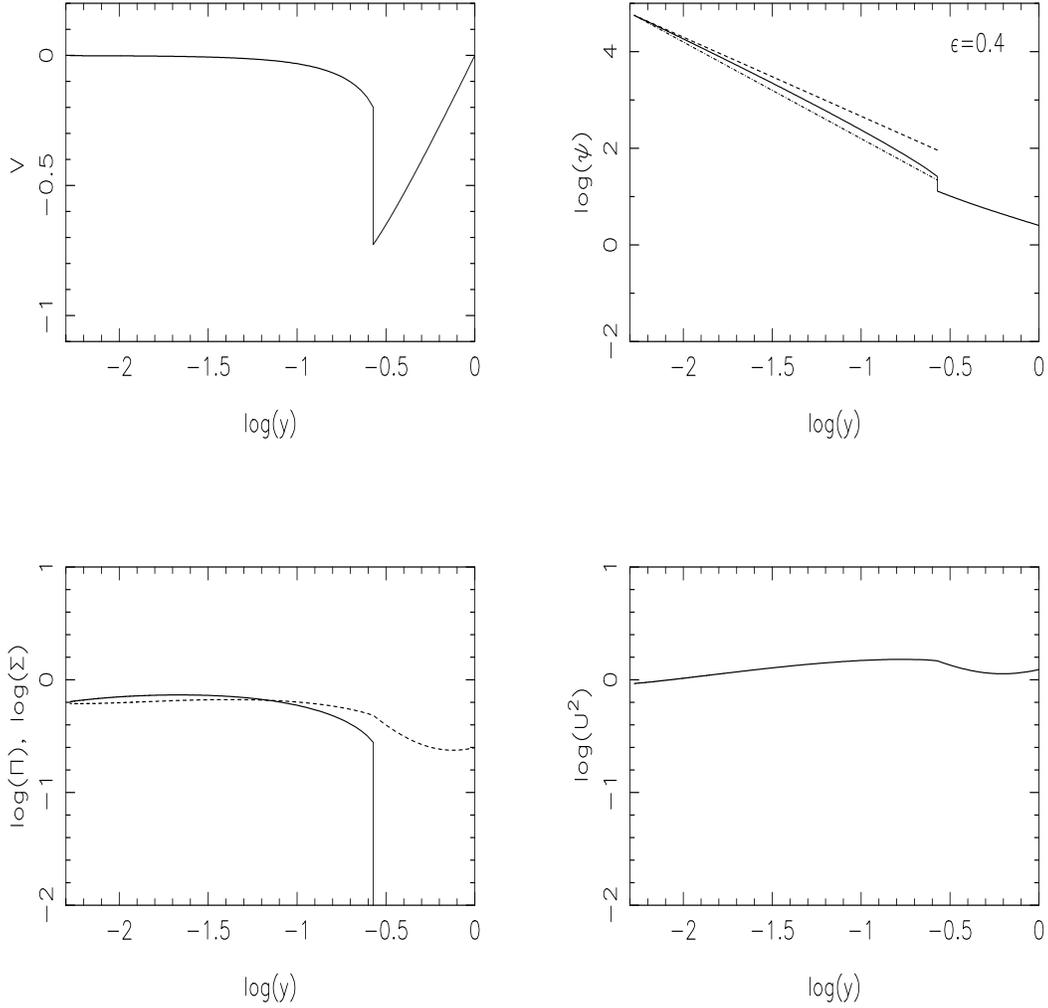}
\end{picture}
\caption{Self similar collapse solution for $\epsilon=0.4$, 
$\tilde{\Sigma}_2(1) = 0.25$, $\tilde{\Pi}(1) =0$. 
Here the tangential velocity dispersions have been introduced 
at the turn around radius corresponding to $y=1$.  The force due to
the tangential velocity dispersion at turn around is $13.5\%$ of the radial
gravitational force.The various quantities shown are same as
in Fig. 2.     
}
\end{figure*}

\clearpage

\begin{figure*}
\centering
\begin{picture}(400,250)
\psfig{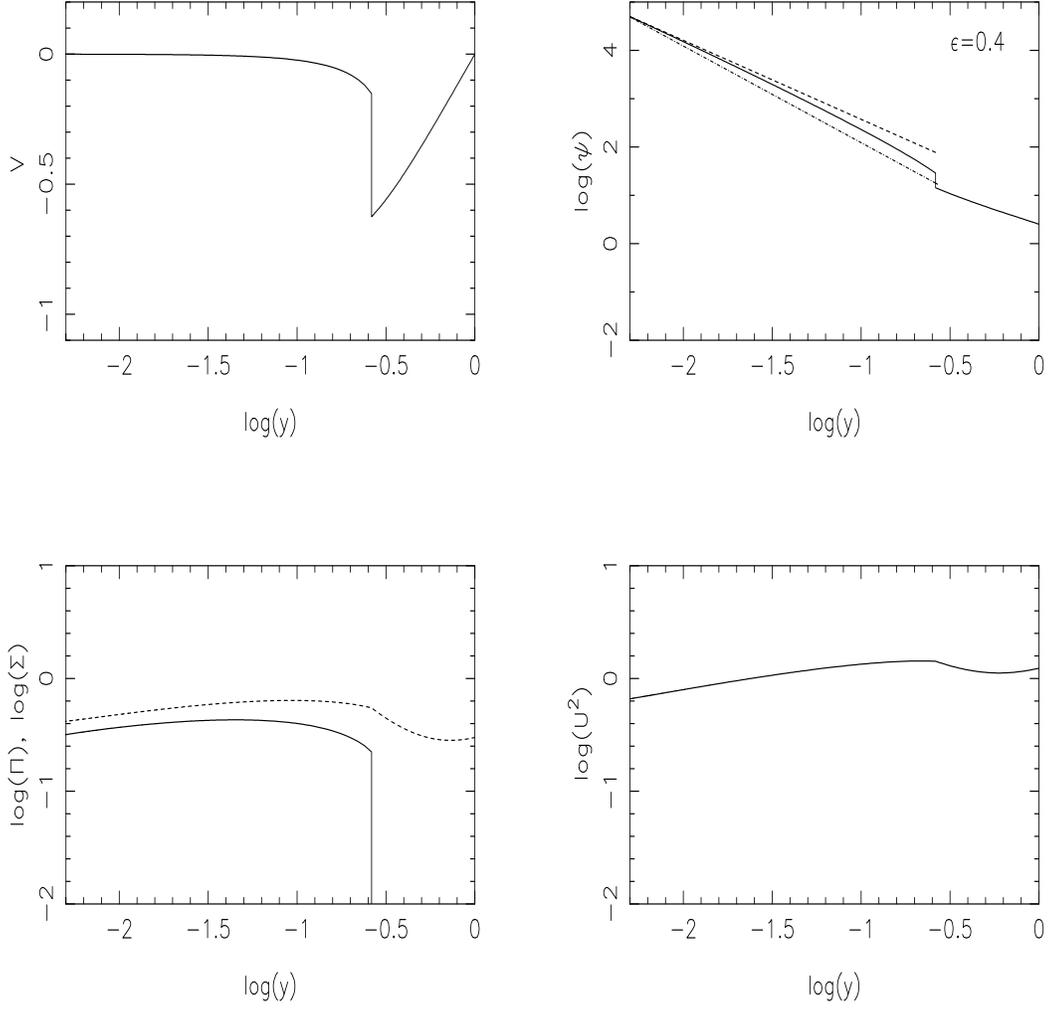}
\end{picture}
\caption{Self similar collapse solution for $\epsilon=0.4$, 
$\tilde{\Sigma}_2(1) = 0.30$, $\tilde{\Pi}(1) =0$. 
The tangential velocity dispersions have been introduced 
at the turn around radius corresponding to $y=1$.
 The various quantities shown are same as in Fig. 2. 
 }
\end{figure*}

\clearpage

\begin{deluxetable}{crrrrrrrrrrrr}
\footnotesize
\tablecaption{The parameters of the self similar solutions \label{tbl-1}}
\tablewidth{0pt}
\tablehead{
\colhead{$\epsilon$} & \colhead{$\tilde{\Sigma}_2$}   & \colhead{$y_s$}   & \colhead{$V_2$} & 
\colhead{$\tilde{\psi}_2$}  &  \colhead{$ \tilde{\Sigma}_2$} 
 & 
\colhead{$\tilde{\Pi}_2$} & 
\colhead{$y_m$}     & \colhead{$V_m$}  & 
\colhead{$\tilde{\psi}(y_m)$}   & \colhead{$ \tilde{\Sigma}(y_m)$} 
& \colhead{$\tilde{\Pi}(y_m)$}          &
}
\startdata
1 &0 & 0.4628 & -0.313 &1.009 &0 & 0.433 &1.7E-4&-4.0E-5& 3.139 & 0&
5.464 \nl
0.4 & 0.94 & 0.4955& -0.0273 & 3.141 &0.94 & 0.517 & 2.86E-3& 
-1.0E-6 &3.641 & 1.126& 0.676  \nl
0.4 & 0.65 &0.3797 & -0.223 & 2.750 & 0.65 & 0.672 & 4.62E-3 & 
-1.75E-3 & 9.691 & 4.661 & 6.162  \nl
1/6 & 2.2 & 0.2584 & -0.0473 & 6.785 & 2.2 &1.232 & 1.12E-2
&-2.0E-3 & 10.16  & 4.021 & 2.261  \nl
1/6 & 2.0 & 0.1999 & -0.163 & 7.140 & 2.0 & 1.584 & 1.56E-2 & -1.5E-2 &
24.17 & 13.45 & 9.615 \nl
& $\tilde{\Sigma}(1)$ & & & & & & & & & & \nl
0.4 & 0.25 & 0.2684 & -0.1995 & 3.024 & 0.78 & 0.449 & 5.2E-3 & -1.27E-3 &
10.67 & 4.180 & 4.318 \nl
0.4 & 0.30 & 0.2622 & -0.1521 & 3.210 & 0.89 & 0.363 & 4.5E-3 & -4.5E-4 
& 8.657 & 2.909 & 2.203 \nl
\enddata


\end{deluxetable}

\begin{deluxetable}{crrrrr}
\footnotesize
\tablecaption{The self similar solutions with $\epsilon=0.4$, 
$\tilde{\Sigma}_2=0.94$ \label{tbl-2}}
\tablewidth{0pt}
\tablehead{
\colhead{$y$} & \colhead{$V(y)$} & 
\colhead{$\tilde{\psi}(y)$}  &  \colhead{$ \tilde{\Sigma}(y)$}  & 
\colhead{$\tilde{\Pi}(y)$} & 
\colhead{$\tilde{M}(y)$} 
}
\startdata
0.4955 &  -2.725E-02 &  3.141   &  0.940 &  0.517 &  30.24 \nl
0.4562 & -2.353E-02 & 3.164     & 0.948 & 0.524 &  30.39\nl
0.4226 &  -2.064E-02 & 3.184    & 0.955 & 0.530  &30.51\nl
0.3937 & -1.832E-02 & 3.201 &  0.961 & 0.535 & 30.62\nl
0.3684 & -1.643E-02 &  3.217 & 0.966 & 0.540  & 30.72\nl
0.3265 & -1.352E-02 & 3.243  & 0.975 & 0.548  & 30.90\nl 
0.2932 &  -1.141E-02 &  3.264 & 0.983& 0.555  & 31.04 \nl
0.2660 & -9.811E-03 &  3.283 & 0.990 & 0.560  & 31.16 \nl
0.2336 & -8.036E-03 &  3.306 & 0.999 &  0.567 &  31.32 \nl
0.2009 & -6.398E-03&  3.331 & 1.008 & 0.575 &  31.49 \nl
0.1710 & -5.027E-03& 3.355 & 1.018 & 0.583 & 31.66 \nl
0.1415 & -3.798E-03 & 3.381 & 1.028 & 0.591 &  31.85 \nl
0.1113 & -2.671E-03  & 3.412 & 1.040 & 0.601 &   32.06 \nl
0.0802 & -1.658E-03 & 3.448 & 1.054 & 0.613  & 32.32 \nl
0.0506 & -8.514E-04 & 3.492 & 1.072 &  0.626 & 32.62 \nl
0.0201 & -2.262E-04 & 3.558 & 1.098 &  0.648  & 33.09 \nl
0.0100 & -8.091E-05 & 3.593 & 1.113 & 0.660   & 33.33 \nl
0.0080 & -5.759E-05 & 3.603 & 1.116 & 0.663  & 33.39 \nl
0.0050 & -2.493E-05 & 3.621 & 1.122 &  0.669 &   33.50 \nl
0.0040 & -1.384E-05 & 3.629 & 1.124 &  0.672 & 33.54 \nl
0.0032 & -4.482E-06 & 3.637 & 1.126 &  0.675 & 33.56 \nl
0.0029 & -1.000E-06 & 3.641 & 1.126 & 0.676  & 33.56 \nl
\enddata

\end{deluxetable}

\clearpage

\end{document}